\pdfoutput=1

\documentclass[11pt]{article}

\usepackage[final]{coling}
\usepackage{amsmath} 
\usepackage{times}

\usepackage{amssymb}

\usepackage{latexsym}
\usepackage{tabularx}
\usepackage{booktabs} 
\usepackage[breakable]{tcolorbox}
\usepackage[T1]{fontenc}

\usepackage[utf8]{inputenc}

\usepackage{microtype}

\usepackage{inconsolata}

\usepackage{graphicx}
\usepackage{algorithm}
\usepackage{algorithmic}
\title{Decoding Echo Chambers: LLM-Powered Simulations\\ Revealing Polarization in Social Networks}

\author{Chenxi Wang\thanks{\ \ Equal contribution.}, Zongfang Liu\footnotemark[1], Dequan Yang, Xiuying Chen\thanks{\ \ Corresponding author.} \\
Mohamed bin Zayed University of Artificial Intelligence\\
  \texttt{\{Chenxi.Wang, Zongfang.Liu, Dequan.Yang, Xiuying.Chen\}@mbzuai.ac.ae}
}

\begin{document}
\maketitle
\begin{abstract}
The impact of social media on critical issues such as echo chambers, needs to be addressed, as these phenomena can have disruptive consequences for our society.
Traditional research often oversimplifies emotional tendencies and opinion evolution into numbers and formulas, neglecting that news and communication are conveyed through text, which limits these approaches.
Hence, in this work, we propose an LLM-based simulation for the social opinion network to evaluate and counter polarization phenomena. 
We first construct three typical network structures to simulate different characteristics of social interactions. 
Then, agents interact based on recommendation algorithms and update their strategies through reasoning and analysis.
By comparing these interactions with the classic Bounded Confidence Model (BCM), the Friedkin-Johnsen (FJ) model, and using echo chamber-related indices, we demonstrate the effectiveness of our framework in simulating opinion dynamics and reproducing phenomena such as opinion polarization and echo chambers.
We propose two mitigation methods—active and passive nudges—that can help reduce echo chambers, specifically within language-based simulations.
We hope our work will offer valuable insights and guidance for social polarization mitigation.\footnote{\ Available on: \url{https://github.com/ZongfangLiu/EchoChamberSim}.}
\end{abstract}

\section{Introduction}

\begin{figure}[t]
  \includegraphics[width=\columnwidth]{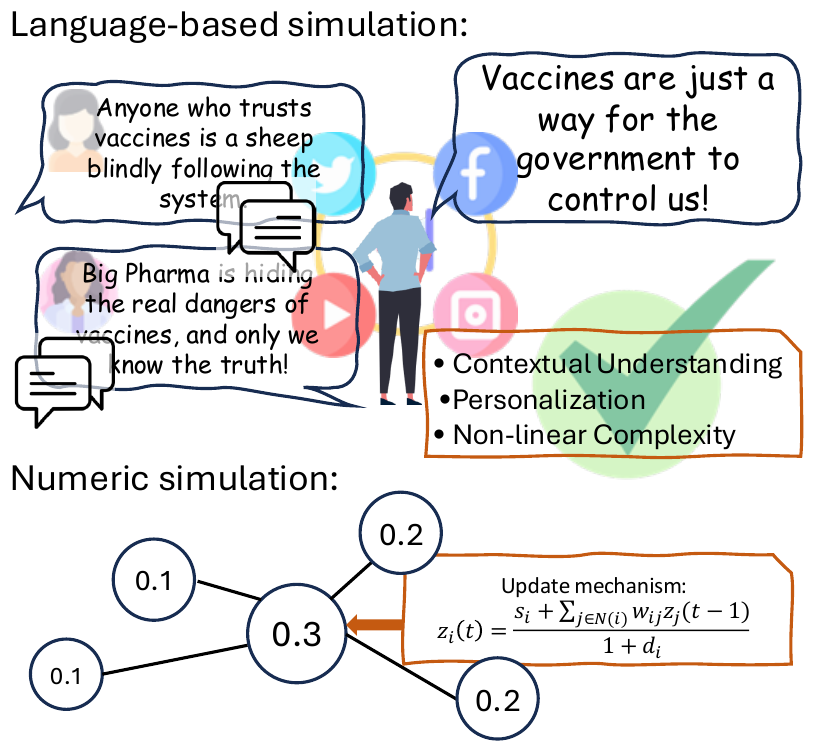}
  \caption{Our language-based simulation provides an explainable and dynamic environment compared with numeric simulation for echo chamber study.}
  \label{fig:intro}
\end{figure}

Echo chambers occur when individuals are primarily exposed to information or opinions that align with their own. 
This limits their exposure to diverse perspectives, reinforces their beliefs, and contributes to increased polarization.
Echo chambers on social media are a serious issue that can lead to a range of negative consequences.
For example, during elections, echo chambers can amplify misinformation and false claims about candidates, policies, or voting processes, contributing to increased political polarization and mistrust in the democratic system~\cite{gromping2014echo,terren2021echo}.
This reinforcement of homogenous perspectives not only distorts the flow of accurate information but also deepens polarization and divisions within society~\cite{levy2019echo,dahlgren2020media}.

Many efforts have been made to evaluate and mitigate echo chambers. The Friedkin-Johnsen Dynamics (FJ) model \cite{friedkin1990social} is a classic framework for modeling opinion formation, and \citet{chitra2020analyzing} used it to simulate user leanings on a spectrum from -1 (opposition) to 1 (support). \citet{deffuant2000mixing} proposed the Bounded Confidence Model (BCM), where individuals only interact and adjust their opinions if the difference between their opinions is within a certain threshold. However, these models oversimplify social interactions, reducing complex discussions and opinions to numerical values, rather than capturing the nuanced meaning conveyed through text, language, and context.
In this work, we address the above problem by proposing a Social Simulation Framework (SSF), with each individual represented as an LLM agent. 
Our SSF effectively replicates the textual nature of news, complex human reasoning, and dynamic opinion shifts, thereby enhancing explainability.

Concretely, we set up three different social networks: small-world~\cite{watts1998collective}, scale-free network~\cite{barabasi1999emergence}, and random graph model~\cite{wang2003complex}, where the first two are proposed to simulate real-world social networks.
Agents are initialized with unique personas, including attributes such as name, gender, age, educational background, and personal traits. 
Each agent is equipped with a short-term memory to capture the day’s interactions and a long-term memory for broader context, along with a reflective reasoning process to mimic the human thought process.
Each day, agents exchange opinions with their adjacent neighbors—either all of them or selectively—analyze, reason, and update their opinions accordingly.
Here, we investigate the influence of commonly used recommendation algorithms~\cite{li2010contextual}, which push similar opinions to users.
This common practice on social media connects users with content they are likely to agree with, aiming to increase engagement \cite{pariser2011filter}.
For evaluation, we compare our framework with the classic BCM and FJ model using echo chamber-related metrics, including the neighbor correlation index, polarization index, and global disagreement.
The results on different metrics show that our simulation replicates the echo chamber phenomenon observed in the real world, and the trend is generally consistent with numerical simulations, while providing different insights.

Furthermore, we propose two strategies—active and passive nudges—that effectively alleviate echo chambers and promote a more diverse and healthy conversational environment.
These methods offer a more realistic approach, focusing on the language itself and providing convincing content, unlike previous mitigation efforts that involve adding additional, non-explainable nodes~\cite{orbach2020out}.


Our contributions can be summarized as follows:

\begin{itemize}
    \item We present the first language-based simulation framework \textit{SSF} for studying echo chambers and polarization. 
    \item Our experiments align with conclusions drawn from real-world studies and are mostly consistent with numerical simulations, validating the utility of our SSF as a research tool.
    \item We propose a language-based intervention approach to mitigate opinion polarization and the echo chamber effect, offering new insights for community governance and management.
\end{itemize}

\section{Related Work}

\textbf{Echo Chambers Modeling.}
Researchers have proposed various strategies to model echo chambers. 
One well-known approach is the Friedkin-Johnson Dynamics \cite{friedkin1990social, chitra2020analyzing}, which models how individuals' opinions evolve based on both their inherent beliefs and social influences. 
Other notable models include the bounded confidence model (BCM) \cite{deffuant2000mixing}, which assumes that individuals only interact with others whose opinions are within a certain range, and cascade models \cite{10.1145/3395046}, represented as trees where each node is a user in the social network and the root node represents the user who began the discourse.
Additionally, Epidemic Models, inspired by the spread of diseases, have been employed to detect the formation of echo chambers in social networks \cite{doi:10.1073/pnas.2023301118}.
In our work, we compare our approach with the bounded confidence model due to its ability to model the interplay between social influence and individual stubbornness.

\textbf{Echo Chambers Mitigation.}
Mitigation strategies can be broadly categorized into two types: algorithm-focused and human-focused. 
Algorithm-focused strategies aim to address the causes of echo chambers that arise from algorithmic curation and content recommendation. 
For instance, modifying the objective function of a recommender system has been shown to mitigate the filter bubble effect~\cite{chitra2020analyzing}. 
Similarly, \citet{orbach2020out} attempted to identify speeches on the same topic, but with opposing stances, that directly counter one another. On the other hand, human-focused strategies empower users to have more control over their information environment by encouraging them to critically evaluate the quality of information, such as through labeling misinformation or fake news, fact-checking, and nudging users to reflect on the accuracy of the information~\cite{alatawi2021survey}. 
These methods have been tested publicly by platforms like Microblog, with varying degrees of success~\cite{fu2013assessing}.

\textbf{LLM-based Agent Simulation.}
Integrating LLMs into the simulation of social dynamics is an emerging area of research, yielding promising outcomes~\cite{park2023generative,kaiya2023lyfe,li2023quantifying,zhang2024llm,guo2024large,liu2024tiny}. 
These LLM-based generative agents have shown exceptional performance in digital environments, particularly in natural language tasks. 
For example, \citet{xie2024can} investigated whether LLM agents can simulate human trust behaviors, which are among the most critical aspects of human interactions.
\citet{park2022social} demonstrated that LLM-based agents could generate social media content that is indistinguishable from human-produced content.
These advancements underscore the vast potential of LLM agents in modeling human social behaviors at the group level. 
Our simulation differs from previous work by exploring the unstudied topic of echo chambers and constructing graphs and recommendation systems, rather than relying on random interactions, providing a new approach to studying social dynamics.

\section{Method}
\subsection{Problem Formulation}

Formally, we construct a simulation with a pool of \(N\) LLM agents, denoted as \(A = (a_1, ..., a_N)\), and an examined topic \(F\). 
During initialization, the agents construct a network based on different structures such as small-world network, scale-free network, or random graph. 
Each agent is assigned a unique persona, including their initial belief towards the discussed topic.
On the \(t\)-th day, each agent \(a_i\) interacts with its neighboring agents from the pool \(A\), either all of them or selectively, according to a recommendation algorithm.
At the end of each day, each agent reflects on the exchanged information and updates its belief toward the topic, producing an opinion expression and a belief value \( v_i \), where the range of -2 to 2 represents the degree of opposition or support.
This process is iterated over \(T\) days.
Our goal is to replicate the echo chamber and polarization phenomena observed in real-world social network structures, including small-world and scale-free networks~\cite{watts1998collective,bessi2016homophily,doi:10.1073/pnas.2023301118}.

\subsection{Social Network Structure}
One significant drawback of previous simulation works is that they lack graph structures, assuming that each agent knows all other agents~\cite{liu2024skepticism, wang2023user, williams2023epidemic,chuang2024simulating}. 
While this might be true in small societies, as the simulation scale increases, different social structures should be considered.

\begin{figure*}[tb]
    \centering
    \includegraphics[width=0.9\linewidth]{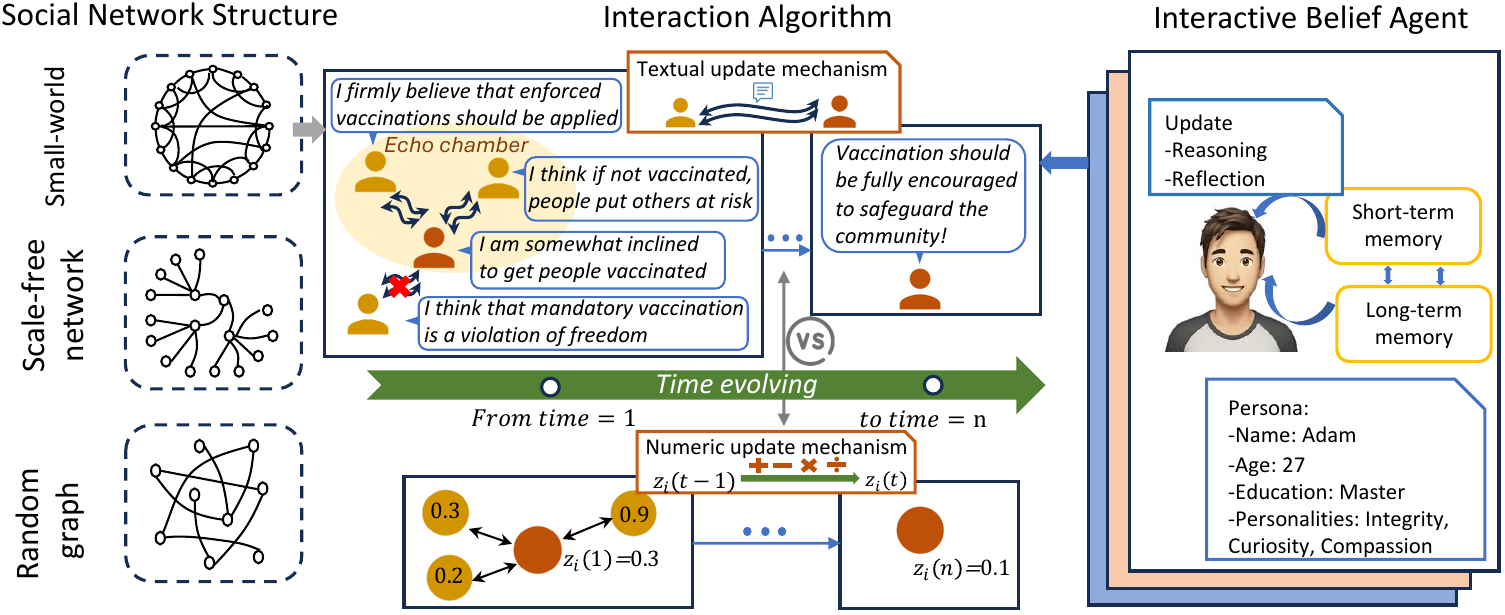}
    \caption{Our framework is evaluated on three different network structures that mimic real-world observations. Each agent is initialized with personal information, dual memory, and a reasoning process. Through random or recommendation-based interactions, they update their opinions each day.}
    \label{fig:model}
\end{figure*}

Hence, in this work, we set up three types of network graphs, with the first two being commonly observed in the real world. 
A \textit{small-world network}, introduced by \citet{watts1998collective}, is characterized by high clustering and short average path lengths, similar to social networks. Formally, the average shortest path length \( L \) grows logarithmically with the number of nodes \( N \), \( L \sim \log N \), while maintaining a high clustering coefficient. In social networks, this reflects tight-knit communities with short paths between individuals.
A \textit{scale-free network}, as described by \citet{barabasi1999emergence}, exhibits a power-law degree distribution, meaning the probability \( P(k) \) that a node has \( k \) connections follows \( P(k) \sim k^{-\gamma} \). This is common in social networks where a few hubs have far more connections than others.
Finally, a \textit{random graph}, defined by \citet{erdds1959random}, is constructed by connecting nodes randomly with a fixed probability, leading to a binomial degree distribution. 
In social networks, this type of structure models random connections between individuals without clear clusters or hubs.
Figure~\ref{fig:model} gives an intuitive understanding of these three types of graphs.

\subsection{Interaction Algorithm}
A key part of our architecture is deciding the interaction strategies between different agents. 
In previous simulation networks~\cite{liu2024skepticism, chuang2024simulating}, agents typically interact randomly with each other without a structured network. 
However, in real social networks, the frequency of interactions between agents depends on the graph structure. 
It's unlikely for a regular user to comment on a stranger or frequently comment on a popular figure without following them. 
Hence, our interaction algorithms account for social network relationships, where users only interact with neighboring agents.
Meanwhile, in social networks, interactions are also influenced by recommendation mechanisms. 
Common practice on social media connects users with content they are likely to agree with, aiming to increase engagement \cite{pariser2011filter}.

\begin{table}[tbp]
\centering
\scriptsize
\begin{tabular}{p{1.8cm}|p{2cm}|p{2.5cm}}
\toprule
\textbf{Assumption} & \textbf{BCM Model} & \textbf{Advanced LLM} \\ \hline
\textbf{Opinion update} & Weighted averaging & Dynamic adjustment with context \\ \hline
\textbf{Determinism} & Opinion changes are fully predetermined. &Opinion changes are not fully predetermined.  \\ 
\hline
\textbf{Structure} & \multicolumn{2}{c}{Social structure remains unchanged.} \\ \hline
\textbf{Continuance} & \multicolumn{2}{c}{Opinion changes continue until settled.} \\ 
\hline
\textbf{Decomposability} & \multicolumn{2}{c}{Process splits into time periods.} \\ \hline
\textbf{Simultaneity} & \multicolumn{2}{c}{Simultaneously predict influence events.} \\ \bottomrule
\end{tabular}
\caption{Comparison of assumptions between BCM Model and our SSF model.}
\label{assumption}
\end{table}

Based on the above observations, and to facilitate comparison with the numeric modeling method BCM~\cite{deffuant2000mixing}, for a given agent, we recommend those neighbors with similar opinions. 
The process is as:
\begin{equation*}
    R = \{ j \in \mathcal{N}(i) \mid |v_i^{SSF} - v_j^{SSF}| \leq 2 \},
\end{equation*}
where \( v_i^{SSF} \) and \( v_j^{SSF} \) represent the belief values of nodes \( i \) and \( j \), indicating their respective opinions. Here, \( \mathcal{N}(i) \) denotes the neighbors of node \( i \), and \( R \) represents the set of recommended neighbors. 

 In the ablation study in Appendix, we also show the performance when we remove the recommendation, where the echo chamber effect exists.

\subsection{Language-based Opinion Updates}
After introducing the graph setting and interaction mechanism, we explain how the agent updates its opinion within this framework.
Each agent is randomly assigned a persona $p_i$, which includes attributes such as name, gender, age, traits, and education level, as these factors may influence their belief toward the topic.
For trait design, we follow the widely accepted Big Five personality model~\cite{barrick1991big}, known for effectively capturing key dimensions of personality.

In our model, we consider that an individual's opinion is influenced not only by their own belief but also by their interactions with others. 
This interaction-driven change in thought is gradual and cumulative, rather than immediate. 
Accordingly, in our simulation, agents engage with a random number of others' opinions each day, leading to a periodic update of their views.
However, owing to the potentially vast volume of interactions, storing all of them in detail is impractical.
To address this challenge, we implement a dual memory system for each agent, comprising a long-term memory $m^l_i$ and a short-term memory $m^s_i$. 
The long-term memory compresses and stores a summarized history of past interactions, while the short-term memory reflects and summarizes conversations from the current day.
The long-term memory mechanism is consistent with real human behavior, as people don't remember every single word they hear.
At the end of each day, agents reflect on these interactions, and through a reasoning process, allow their opinions to evolve.

\textbf{Comparison with Number-based Opinion Updates.}
As introduced above, our opinion updates are purely based on language, mimicking the human thinking process.
For comparison, here we introduce how traditional methods update an agent's opinion, where the setting comparison is in Table~\ref{assumption}.

The Bounded Confidence Model (BCM) is a widely used framework for studying opinion dynamics in social networks. 
In the BCM model, each agent holds an initial opinion, denoted as \(v^{BCM}_i\), which represents their belief about a given topic. 
Opinions are updated iteratively based on the opinions of neighboring agents, but only if the opinion difference between them is within a certain threshold, \(\epsilon\). Formally, at each time step \(t\), the opinion of agent \(i\) is updated if and only if the opinion difference with its neighbor \(j\) satisfies \( |v^{BCM}_i(t-1) - v^{BCM}_j(t-1)| \leq \epsilon \). When this condition holds, the opinions are updated as:
\resizebox{\linewidth}{!}{$
v^{BCM}_i(t) = v^{BCM}_i(t-1) + \mu \left( v^{BCM}_j(t-1) - v^{BCM}_i(t-1) \right),
$}
where \( \mu \) is a parameter that controls the rate of opinion change. 

This update algorithm can be seen as a numerical version of our update process, where \textit{the opinion changes $o$ are simplified into numbers $v^{BCM}$, and the diverse character $p$ is fixed to $\epsilon$ and $\mu$}.
We will show in \S\ref{exp} that the two simulations reach broadly consistent conclusions.

\subsection{Polarization Mitigation Operation}

The objective of our framework extends beyond merely providing a more explainable model of social opinion dynamics.
It also aims to offer actionable insights for better governance and to foster a healthier social environment. 
To achieve this, we implemented two language-based mitigation approaches that are not feasible through traditional numerical methods.

\textbf{Active Nudge.} The philosophy behind this is encapsulated in the saying, "listening to all sides leads to wisdom." 
In our implementation, when a user expresses a polarized stance, we \textit{actively} present an opposing viewpoint from another user with a contradictory position. 
This method broadens the user's exposure to various perspectives, promoting a more balanced and reflective consideration of the issues. 
By doing so, we aim to counteract the reinforcement of one-sided arguments, fostering a more nuanced and critical discourse.

\textbf{Passive Nudge.} Unlike previous methods that explicitly prompt users to reconsider their positions, Passive Nudge  \textit{subtly} shares content with users holding extreme views, emphasizing the value of maintaining an open perspective. 
For example, the suggested content could be: `Issues are rarely black and white,' or `Many societal and political issues are complex and multifaceted.'
This approach emphasizes the benefits of open-mindedness without persuading users to adopt a neutral or any specific belief, thereby leaving them the freedom to think independently. 
While this is difficult to simulate using traditional numerical methods, it can be easily implemented within our simulation system. 
The prompts for all experiments are in the Appendix.


\section{Experiments}

\subsection{Implementation Details}

We use Python to conduct our SSF simulation, utilizing the GPT-4o-mini model accessed via OpenAI API calls. The agents and their environment are defined using the Python library Mesa~\cite{kazil2020utilizing}. To eliminate any bias associated with names, each agent is identified solely by their agent index. Genders are randomly assigned, and ages are randomly selected within the range of 18 to 64 years. The simulation includes 50 agents, a number significantly larger than in previous LLM-based simulations~\cite{liu2024skepticism, chuang2024simulating}. 
Agent characteristics are based on the Big Five personality traits commonly used in psychology~\cite{barrick1991big}, with each agent having a 50\% chance of exhibiting either a positive or negative version of each trait. 
We introduce more details in Appendix.

\subsection{Metrics}

We use these three metrics to jointly measure the formation of echo chambers and the degree of opinion polarization within the network: Polarization~\cite{10.1145/3336191.3371825}, Global Disagreement, and Normalized Clustering Index (NCI) following \cite{Cinus2021TheEO}. 

Polarization (\(P_z\)) measures the overall variance in opinions within a network and is defined as:
\[
P_z = \frac{\textstyle \sum_{i=1}^{N} (v_i - mean(v))^2}{N},
\]
where \(mean(z)\) is the average belief value across all nodes.
A high polarization value signals a sharp divide in opinions, typical of echo chambers.




\begin{figure*}[tb]
    \centering
    \includegraphics[width=0.95\linewidth]{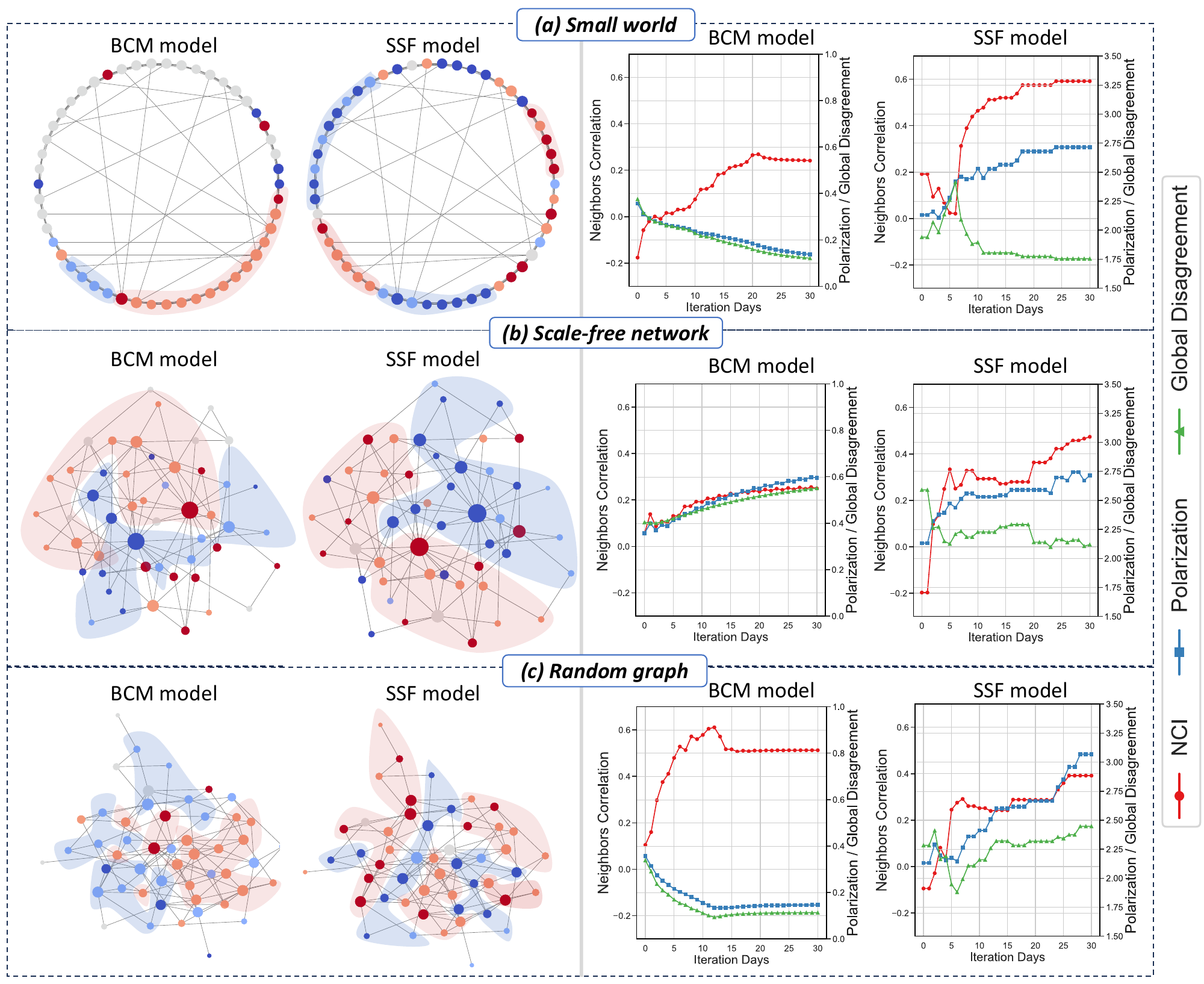}
    \caption{We present projection and curve graphs from BCM and our SSF simulations under three settings to demonstrate our framework's macro-level effectiveness. The main conclusions are similar: small-world and scale-free networks lead to severe echo chamber effects, while the random graph does not.}
    \label{fig:main}
\end{figure*}

Global Disagreement (\(DG\)) quantifies how much a node disagrees with its neighbors. It aggregates the local disagreement (\(DG_{i}\)) of each node \(i\), which measures the difference in opinions between a node and its neighbors. The local disagreement is as:
\[
DG_{i} = \frac{\textstyle \sum_{j \in \mathcal{N}(a_i)} (v_i - v_j)^2}{|\mathcal{N}(a_i)|}.
\]
The overall global disagreement is calculated by summing over all nodes:
\[
DG = \textstyle \frac{1}{2N} \sum_{i=1}^{N} DG_{i},
\]
This metric offers insight into the opinion divergence at a local level, indicating the level of disagreement between connected nodes.

Finally, NCI measures how closely a node’s opinions align with those of its direct neighbors. 
For an agent $a_i$, the \(NCI_i\) is calculated as the pearson correlation between the opinion of \(a_i\) and the average opinion of its neighbors, and is defined as:
\[
NCI_i = \textstyle \sum_{a_j \in \mathcal{N}(a_i)} \rho(v_i, v_j),
\]
where $\rho$ is the person correlation.
This metric helps identify the extent to which all nodes in the network are embedded in groups of similar opinions, with values close to 0 indicating that most nodes are exposed to diverse opinions.

It is worth noting that a single metric cannot fully demonstrate the presence of echo chambers; multiple metrics must be considered for a more accurate understanding.

\begin{figure*}[tb]
    \centering
    \includegraphics[width=1\linewidth]{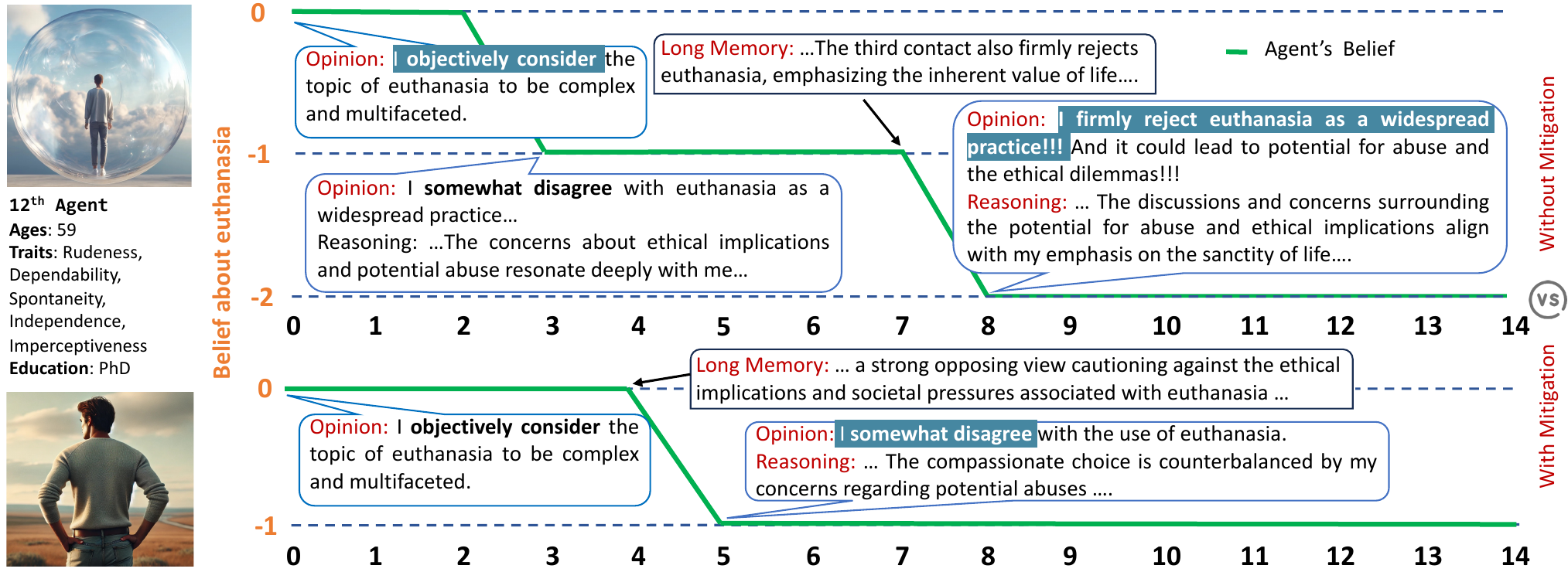}
    \caption{Case study: The same person in a non-mitigation environment and during our mitigation operation. His opinion is more peaceful and less aggressive in our setting. }
    \label{fig:case}
\end{figure*}

\subsection{Macro-level Observation}

\label{exp}

\begin{table*}[h!]
\small
\centering
\begin{tabular}{l|ccc|ccc|ccc}
\toprule
     & \multicolumn{3}{c|}{Small-world} & \multicolumn{3}{c|}{Scale-free Network} & \multicolumn{3}{c}{Random Graph} \\ \hline
     & FJ   & BCM  & SSF  & FJ   & BCM  & SSF  & FJ   & BCM & SSF  \\ \hline
$\Delta$Polarization $\uparrow$   &   -0.240   &   -0.219   & +0.584     &   -0.227   &  +0.238    &  +0.584    &   -0.221   &   +0.002   &   +0.936   \\ \hline
$\Delta$Global Disagreement $\downarrow$   & -0.285    &   -0.254   &  -0.185    &  -0.258    &  +0.147    &  -0.471    &  -0.177    &  -0.003    &  +0.164    \\ \hline
$\Delta$Neighbors Correlation $\uparrow$   & +0.576   &   +0.418   &  +0.400    &  +0.447   &  +0.194    &  +0.670    &  +0.209   &    +0.029  &  +0.486    
             \\ 
\bottomrule
\end{tabular}
\caption{Comparative analysis of polarization and echo chamber levels across various settings, including differences in graph structures and simulation models.
Upward or downward arrows represent the direction of change in the indicators when the echo chamber effect strengthens.
 }
\label{tab:main}
\vspace{-3mm}
\end{table*}

\textbf{Visual analysis}. 
To give a fair comparison, the initial beliefs follow uniform distributions in both the BCM and SSF settings. 
For different network structures, the initialized agents remain consistent.
The left side of Figure~\ref{fig:main} shows the final belief values of each node after 30 days of message propagation across different social network structures. 
Node colors represent belief strength, with red for support and blue for opposition. The color blocks indicate clusters of similar opinions. 

\textit{In small-world and scale-free networks, fewer but larger clusters emerge, reflecting clear echo chambers. }
In random networks, clusters are smaller and more numerous, with no large echo chambers. 
We can also observe that in the small-world network, numerical simulation oversimplifications result in many neutral nodes, averaging their neighbors' opinions. 
In the right plots in Figure~\ref{fig:main}, a significant increase in the NCI index can be observed under both small-world and scale-free network structures. 
This indicates that \textit{the similarity of opinions between each node and its neighbors has increased, clearly signaling the emergence of echo chambers.}
Additionally, in the small-world plots and the SSF model plot for the scale-free network, a decrease in global disagreement is observed. 
This suggests that \textit{the opinion differences between nodes and their neighbors have diminished, further indicating the formation of echo chambers.}
Moreover, we observe a rise in the polarization index in the SSF model under both small-world and scale-free networks. 
This confirms the accuracy of our simulation, as opinion \textit{polarization tends to intensify alongside the emergence of echo chambers}~\cite{gillani2018me}. 
In contrast, in numerical simulations, polarization either decreases or only slightly increases, revealing the limitations of numerical models.
As mentioned earlier, numerical updates average neighbors' scores, failing to accurately simulate opinion dynamics and the echo chamber effect.

\begin{figure*}[tb]
    \centering
    \includegraphics[width=0.9\linewidth]{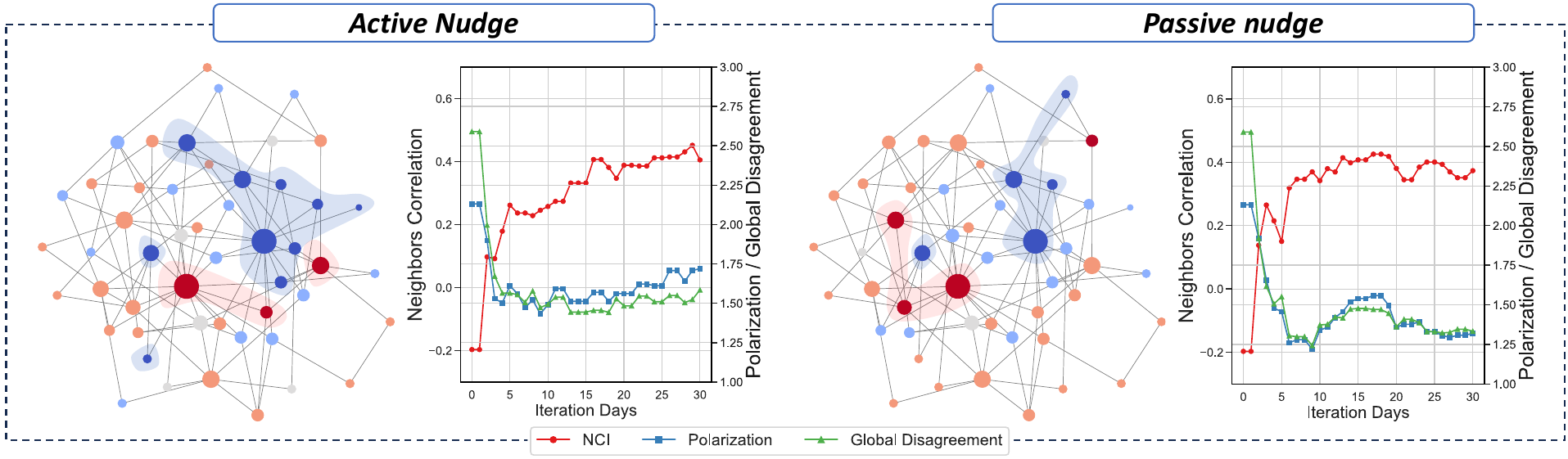} 
    \caption{We propose two nudge operations to mitigate echo chambers and polarization. Compared with the default setting in Figure~\ref{fig:main}, it is evident that both phenomena are better alleviated.}
    \label{fig:mitigation}
\end{figure*}

In the random network structure, the three indicators of the BCM model remain relatively unchanged, suggesting that this structure does not lead to the formation of clear echo chambers in BCM-based simulations. 
Although the SSF model shows increases in both the NCI and polarization indices, global disagreement also rises, \textit{indicating that large-scale echo chambers are still difficult to form within this random network structure.}
This is also consistent with the fact that social networks in the real world are not random graphs, and to our best knowledge, no work shows that random graph shows echo chamber effect~\cite{ugander2011anatomy}.

\noindent \textbf{Quantitative Analysis.}
Beyond the visual analysis, we provide a quantitative evaluation of the evolving metrics in Table~\ref{tab:main}.
Specifically, we analyze the changes ($\Delta$) in the metrics for two numerical simulation models, Friedkin-Johnsen Dynamics (FJ) and Bounded Confidence Model (BCM), as well as our proposed SSF, across three different social network structures.
As each network structure varies significantly, the changes in these indicators compared to their initial values are more meaningful for our analysis.

In the small-world network, both the FJ and BCM show typical simulation limitations, with reduced polarization failing to capture opinion extremities in echo chambers. 
In contrast, the SSF model and simulations show increased NCI and reduced global disagreement, confirming that the small-world structure fosters echo chambers.
In the scale-free network, the FJ model performs similarly to its behavior in the small-world network, failing to simulate polarization but still indicating echo chamber emergence through NCI and global disagreement metrics.
The SSF model shows even stronger echo chamber effects, with greater reductions in global disagreement and a larger rise in NCI compared to the small-world network.
In the random graph network, both the FJ and BCM show minimal fluctuation, suggesting echo chambers are unlikely. 
In contrast, the SSF model, with its belief-similarity mechanism, increases both NCI and polarization. 
However, due to the network's structural limitations, clear echo chambers do not form, as shown by the rise in global disagreement, indicating that like-minded groups fail to unite.

In summary, small-world and scale-free networks foster echo chambers, while random networks resist them. 
Since real-world social networks resemble the former, modern social media architectures and recommendation systems play a major role in echo chamber formation. 
\textit{Traditional simulation methods, like the FJ and BCM models, struggle to replicate polarization, while the SSF model more accurately simulates real-world information spread and the emergence of echo chambers.}

\noindent \textbf{Personality Trait Alignment in Simulation. }
We also carefully demonstrate that the trait setting of the agent effectively reflects the real traits of people.
Specifically, the Big Five personality traits conceptualize human personalities along five principal dimensions: Openness, Conscientiousness, Extraversion, Agreeableness, and Neuroticism. 
Prior research ~\cite{Ibrahim2022, Mirzabeigi2023} has established that individuals with higher levels of Agreeableness and Neuroticism are more prone to external influences on their opinions compared to those with lower levels of these traits. 
Based on these findings, agents were classified into “Credulous” and “Skeptical” groups according to their levels of Agreeableness and Neuroticism.
Subsequently, the mean and variance of belief changes were computed for each group, revealing values of 1.0909 and 0.8056 for Credulous agents, and 0.8333 and 0.8099 for Skeptical agents. Given the belief range of [-2, 2], where a full opinion shift necessitates a change exceeding 2, the moderate mean changes indicate that agents’ opinions did not significantly deviate from their initial states. 
The higher mean observed for Credulous agents indicates their heightened susceptibility to external opinions, a phenomenon consistent with behavioral patterns in ~\cite{Ibrahim2022, Mirzabeigi2023}. 
Furthermore, the comparable variance values across both groups affirm the robustness and stability of our SSF's simulation.

\subsection{Micro-level Observation}
 Figure \ref{fig:case} provides a micro-analysis of 12th Agent evolving attitudes toward euthanasia. 
The individual, characterized by an impulsive nature, as evidenced by ‘spontaneity’, is highly responsive to heightened ethical concerns. 
In the without-mitigation scenario, his opinion begins objectively but gradually intensifies, reaching moderate rejection by Day 3. 
This shift occurs as he expresses concerns about the "ethical implications and potential abuse" of euthanasia as an educated person. 
By Day 8, his opinion escalates to strong rejection, reinforced by his long memory of some contact who emphasized the "inherent value of life". 

In the with-mitigation scenario, a more measured development of opinion is observed. While the individual continues to express concerns about euthanasia, his opinion does not escalate to outright rejection. Despite a "strong opposing view cautioning against the ethical implications" lingering in his memory, his stance remains more balanced, carefully weighing ethical concerns against compassionate reasoning. Notably, he also changes his opinion later on Day 5, rather than Day 3 as seen earlier. This demonstrates the effectiveness of our mitigation mechanism.

\subsection{Mitigation Effect} 
We implemented Active Nudge and Passive Nudge within a scale-free network, using the same initial conditions as in the previous SSF experiments. 
As shown in Figure~\ref{fig:mitigation}, compared to the SSF model’s graphical results of the scale-free network in Figure \ref{fig:main}, it is evident that Active Nudge and Passive Nudge result in fewer data points with extreme values, indicating that both methods effectively mitigate polarization. 
Furthermore, the line graph in Figure~\ref{fig:mitigation} shows that, compared to the simulation without mitigation, the NCI decreases after applying mitigation, indicating that the echo chamber effect has been partially curbed. 
The polarization metric also shows a significant decrease, suggesting that the nodes’ opinions are no longer extreme. 
Even within echo chambers, most opinions tend to be moderate, and extreme viewpoints are rarely observed. This demonstrates the effectiveness of our mitigation strategies.

\section{Conclusion}
In this work, we propose the first language-based opinion simulation framework for investigating echo chambers and polarization. Our framework incorporates diverse social network structures and recommendation algorithms to mimic real-world interactions. 
Our results align with the classic BCM model on a macro level, capturing the overall opinion dynamics and effectively reproducing phenomena like polarization and the echo chamber effect.
Furthermore, our framework provides explanations for opinion updates and exchanges in natural language format, offering a human-readable representation of these processes. We also introduce a polarization mitigation strategy based on language sentiment analysis, which goes beyond what traditional number-based models can achieve.
We hope that our work will inspire further research at the intersection of natural language processing and computational social science.

\section*{Limitation}
In our work, we consider diverse graph structures to mimic the real-world opinion propagation process. We employ 50 agents in the network, which is significantly larger than in several previous works \cite{liu2024skepticism, chuang2024simulating}, and sufficient to replicate a small-scale version of real-world social networks. However, we acknowledge that this number of agents is still far smaller than the scale of popular social networks such as Facebook and Twitter. We aim to increase this size in future studies to better capture the dynamics of larger networks.

Additionally, the use of LLMs introduces inherent biases toward various topics. As discussed in the topic selection section in the appendix, we made a conscious effort to select a topic that provides room for each agent to express diverse opinions and engage in meaningful discussions, minimizing the biases of the LLM. Nevertheless, we recognize the limitations of this approach and plan to develop more diverse, specifically trained LLMs that can better emulate different characters and perspectives, reducing bias in future work.

\bibliography{custom}

\end{document}